# A Review of and Roadmap for Data Science and Machine Learning for the Neuropsychiatric Phenotype of Autism

Peter Washington[1] and Dennis P. Wall[2]

[1]Information and Computer Sciences, University of Hawaiʻi at Mānoa
[2]Pediatrics (Systems Medicine), Biomedical Data Science, Psychiatry & Behavioral Sciences, Stanford University School of Medicine

**Abstract**

Autism Spectrum Disorder (autism) is a neurodevelopmental delay which affects at least 1 in 44 children. Like many neurological disorder phenotypes, the diagnostic features are observable, can be tracked over time, and can be managed or even eliminated through proper therapy and treatments. Yet, there are major bottlenecks in the diagnostic, therapeutic, and longitudinal tracking pipelines for autism and related delays, creating an opportunity for novel data science solutions to augment and transform existing workflows and provide access to services for more affected families. Several prior efforts conducted by a multitude of research labs have spawned great progress towards improved digital diagnostics and digital therapies for children with autism. We review the literature of digital health methods for autism behavior quantification using data science. We describe both case-control studies and classification systems for digital phenotyping. We then discuss digital diagnostics and therapeutics which integrate machine learning models of autism-related behaviors, including the factors which must be addressed for translational use. Finally, we describe ongoing challenges and potent opportunities for the field of autism data science. Given the heterogeneous nature of autism and the complexities of the relevant behaviors, this review contains insights which are relevant to neurological behavior analysis and digital psychiatry more broadly.

**Introduction**

Autism Spectrum Disorder (autism) is a complex neuropsychiatric condition which manifests in a variety of phenotypic presentations. These include limited ranges of interest, social deficits, delays in communication, inability to express and or recognize ranges of emotion, avoidance of eye contact, and idiosyncratic motions of hands, head, and body (Figure 1).

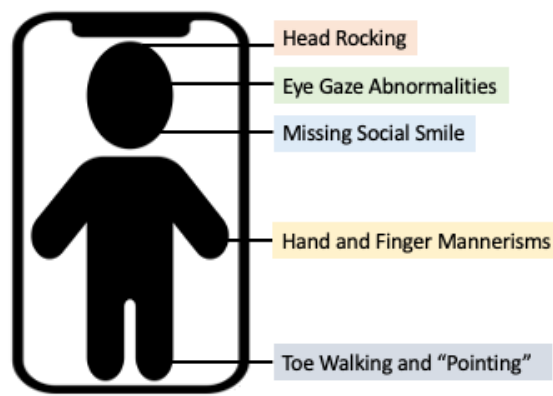

**Figure 1**. Examples of some of the observable features indicative of autism in a developing child 2 years of age. These features can be captured through traditional clinical techniques that land in structured electronic medical records and through alternative methods such as through games on smartphones. The more passively such data can be collected [3] via increasingly ubiquitous devices, the higher the chances of capturing natural signals that have higher diagnostic precision. This is in contrast to signals captured in the more artificial clinical settings where transiency and other factors could cause misleading readings or inaccurate measurements.

It is estimated that 1 In 44 children in the United States have autism [1], with the prevalence continuing to rise. Despite the high incidence rate, it is estimated that 83.86% of counties in the United States lack access to diagnostic and therapeutic resources [2]. This dearth of resources leads to diagnoses later in life, often resulting in poorer psychiatric outcomes [4-5].

Artificial intelligence (AI), and particularly machine learning (ML), has the potential to serve as the great equalizer for many behavioral healthcare concerns like autism. According to Pew Research Center, 97% of adults in the United States own a cellular device and 85% own a smartphone, with the percentage increasing each year since 2011 [6]. In emerging economies such as Mexico, Venezuela, Colombia, South Africa, Kenya, India, Vietnam, the Philippines, Tunisia, Jordan, and Lebanon, the majority of adults own access to a mobile phone of some kind, with 53% possessing access to the Internet and apps [7]. As these percentages continue to rise and Internet-powered devices become increasingly ubiquitous, access to digital services can become democratized on a global scale. While autism services are currently restricted to relatively privileged populations, digital solutions powered by emerging data science methodologies can make access to autism therapy universal.

This review focuses on digital data modalities which can aid in a quantitative understanding of the neuropsychiatric phenotype. Characterized by several behavioral manifestations, the field of autism behavioral data science is vast and broad. The heterogeneous nature of autism presentation requires quantification of autism across several social and communicative dimensions to both better understand the autism phenotype as it relates to differences in diagnostic cohorts (Figure 2A) and to engineer automated methods for autism classification (Figure 2B).

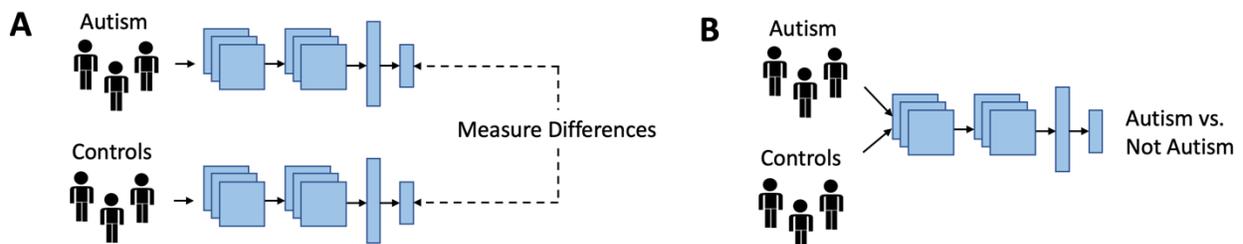

**Figure 2**. The two major classes of studies we review. (A) Case-control studies (science) comparing the output of a static model or computational analysis method between two or more groups, usually autism vs. not autism. (B) Machine learning studies (engineering) with the goal of training a model to classify autism vs. not autism.

The field of autism data science is so vast that it is impossible for us to exhaustively cover the field within the scope of a single review. To maintain focus, we do not cover details of the models and ML techniques powering the studies we describe, though we briefly mention them. The state-of-the-art of ML and data science is constantly evolving, and the details of the exact models used are not as relevant as the content of the data. We make an exception in cases where the model design is crucial to understanding the contribution of the work.

We first describe multimedia data modalities which are commonly used in digital phenotyping projects related to social human behavior, including common schemes for acquiring naturalistic data. We then discuss case-control studies for identifying statistically significant differences between cohorts with and without autism. We next review efforts to automatically classify autism using ML. Finally, we highlight ongoing challenges for the field, corresponding areas of opportunity for innovators to make a significant impact, and "low-hanging fruit" based on successful studies and methods applied to other behaviors and neuropsychiatric presentations.

**Data Modalities and Acquisition**

There are a multitude of data types which encode behavior-rich data for digital phenotyping (Table 1). Data which require humans in the loop to record, such as questionnaires about a child's behavior, are the least prone to computational errors and most comparable to the formal clinical diagnostic process. On the other hand, multimedia data streams obtained by fully automated procedures can comprise of subtle yet complex behavioral indications which would be missed by a human using predetermined prompts. Such multimedia data streams, however, are more difficult to successfully convert into a predictive embedding. Here, we discuss both categories of behavioral data modalities used in the field of autism ML.

| Modality | Data Capture Techniques | Feature Extraction Techniques (Non-Deep Learning) |
|---|---|---|
| Questionnaire/Instrument | Parent-filled questionnaire (on web or smartphone), clinician-filled questionnaire | Conversion to Likert scale |
| Eye gaze | Eye tracker, webcam, video camera | Gaze heat maps, gaze trajectories |
| Emotion evocation | Webcam, video camera | Histogram of Oriented Gradients, facial key points |
| Restrictive and repetitive motions | Webcam, video camera, motion sensor, depth sensor | Skeletal pose, optical flow, activity recognition |
| Idiosyncratic speech | Audio recorder | Many audio-specific features such as pitch, Mel-Frequency Cepstral Coefficients, temporal shape, temporal features, spectral shape, etc. |

**Table 1**. Common data modalities used in both case-control and machine learning autism phenotyping studies. With the continually expanding popularity of deep learning models in behavioral data science (and the entirety of data science more broadly), many features are now automatically learned via neural networks rather than explicitly pre-specified.

One of the most common data sources for clinical phenotyping consists of multiple-choice questionnaires either containing or derived from clinical scoresheets for diagnosing or quantifying the magnitude of the child's presentation of autism. The most popular diagnostic assessments for autism are the Autism Diagnostic Observation Schedule-2 (ADOS) [8], the Autism Diagnostic Interview-Revised (ADI-R) [9], the Childhood Autism Rating Scale (CARS) [10], and the Gilliam Autism Rating Scale-2 (GARS-2) [11]. Other scoresheets are used for screening or quantifying autism-related behaviors. Readily available datasets with such scoresheets filled by clinicians or parents include Autism Genetic Resource Exchange (AGRE) [12], Autism Consortium (AC), National Database for Autism Research (NDAR) [13], Simons Simplex Collection (SSC) [14], Simons Variation in Individuals Project (SVIP) [15], Autism Speaks (MSSNG), and Autism Genome Project (AGP) [16]. Data-driven studies either directly analyze collections of these scoresheets or develop independent questionnaires of a similar nature to these studies. The primary benefits of this data modality are its constrained search space relative to high-dimensional multimedia data and the high degree of reliability of the data due to its requirement of incorporating humans in the feature extraction process.

Rich multimedia streams are increasingly used in data science-based phenotyping efforts. Popular data streams enriched for social human behavior relevant to autism include single images [17], videos [17], eye tracking [17], movement records [17], text [18], and audio [19]. The principal advantage of these data types is the rich and nuanced behavioral information which they encode. The challenge with such large and heterogeneous data, like in many other applications of data science, is finding the signal in the noise.

Most autism data science studies which collect phenotype-rich data utilize controlled lab settings for acquisition. This practice enables normalized data that can be easily compared across subjects. We cover such studies in this review. Recent innovations, however, center around collecting these data in naturalistic settings, and there is extensive room for novelty around the acquisition of homogeneous and structured data streams. Stanford University's *Guess What* project, for example, provides a framework for capturing videos of children with autism enriched for social behaviors through a Charades game [20]. In this game, a parent or caregiver places a smartphone on their forehead while the child acts out Charades prompts. The parent tilts the phone forward when the prompt is correctly guessed and backwards otherwise [21]. The game prompts in conjunction with accelerometer and gyroscope logs of the phone tilting motion provide structured timestamped metadata which can serve as automatically derived labels of emotions, body movements, and other behaviors [22]. Videos collected from *Guess What* sessions with emotion game prompts were demonstrated to be enriched with emotion [23]. The resulting frames were labeled and outperformed general-purpose facial emotion recognition classifiers [24-27]. While *Guess What* was originally developed as a data curation device, feasibility testing has demonstrated improvements on the Social Responsiveness Score-2 (SRS-2) and the Vineland Adaptive Behavior Scales-II (VABS-II) with at-home participation by families with a child with autism [28].

Phenotype-rich data can also be captured from digital therapies which were not originally developed for the purpose of data acquisition. Socially assistive robots have been created to provide therapy through social play and taking the lead in social interactions with the child, thereby evoking social behavior by the child [29]. Structured interactions with robots coupled with measurements from passive sensors can provide objective measurements that can be used for diagnoses [30]. Wearable devices have also been explored as therapeutic tools. The *Superpower Glass* project is an educational tool deployed on a wearable headset to provide real-time facial emotion feedback to children with autism [31-33]. This device provided pleasant experiences to children and their parents [34-35], improved child performance of the SRS-2 [36-37], and ultimately resulted in significant improvements on the VABS socialization subscale in a randomized controlled trial [38]. While video and eye gaze data collected by the *Superpower Glass* system were evaluated for discriminative power in an autism vs. neurotypical classification task, the classifier was unable to outperform classifiers which only used age and gender features [39], likely due to the non-standardization of the interaction between users. Extra care must therefore be taken to ensure that data acquired from interactive systems are calibrated and standardized across users.

As we will see in the following sections, dataset size and diversity are some of the greatest limitations of the current state of autism data science. Innovation in the structured data collection, cleaning, and labeling process are likely to enable well-powered studies and robust diagnostic models.

**Case-Control Studies**

We describe scientific studies which use data science methods to quantify differences between cohorts with autism and matched neurotypical controls. Such experiments aim to gather equivalent data from multiple groups and determine if there are statistically significant differences.

Gaze features are frequently used for case-control studies since avoidance of eye contact is a well-documented autism behavior. Several studies use an eye tracker to measure a child's attention to structured socially enriched video scenes. Jones et al. compared fixation patterns over several months of 59 children at high risk for autism (defined as full siblings of a child with autism) against 51 typically developing children, finding that children later diagnosed with autism exhibit a decline in fixation time from 2 to 6 months of age [40]. Riby et al. compared 26 children with autism and 18 with Williams syndrome, a rare developmental disorder characterized by intellectual disabilities, distinctive facial features, and cardiovascular issues. The children in the study watched cartoon scenes while their gaze was measured. The authors found that children with autism watched faces less time than typical whereas children with Williams attended to the face for a longer duration than is typical [41]. Chawarska et al. compared visual attention to a social scene by 67 children at high risk for autism (full sibling of a child with autism) against 50 children at low risk using eye tracking, finding that infants later diagnosed for autism viewed the scene for a shorter duration and spent less time looking at the actress in the scene and her face [42]. Campbell et al. compared visual attention to a scene for 22 toddlers with autism and 82 neurotypical controls, finding that 8% of participants with autism oriented to name calling compared with 63% of neurotypical participants [43]. Sadria et al. applied network analysis to eye tracking data by considering each area of interest as a node and each saccadic transition between two areas defining an edge. They found that degree centrality yielded statistically significant differences between children with autism and neurotypical controls for areas pertaining to the mouth and right eye [44]. Breaking from the more standard practice of using highly structured interactions and in-lab settings for data collection, Varma et al. measured eye gaze differences between cohorts using network analysis from crowdsourced data collected during use of a mobile autism therapeutic, finding a statistically significant different between the groups for a single area of interest [45]. Alvari et al. analyzed eye contact during unconstrained therapist-child interactions by applying unsupervised clustering on data from 62 children with autism, identifying three distinct subgroups defined by eye contact dynamics [46]. In one of the largest autism-related eye tracking studies containing 563 subjects with autism and 1,300 other subjects, children with autism exhibited the higher percent fixation to dynamic geometric images compared to other children [47]. In an eye tracking study of 21 children with autism and 31 typically developing children playing the Go/No-Go game, Putra et al. found that children with autism had more unstable gaze modulation [48].

Difficulty understanding and evoking emotion is another core difficulty in children with autism. Affective data can therefore be useful in distinguishing autism from neurotypical controls. Guha et al. compared the facial dynamics of 20 participants with high function autism against 19 participants with neurotypical development, finding reduced complexity in the dynamics of the eye region in the autism group [49]. In an uncharacteristically large-scale study for the field, Egger et al. compared emotional features in videos collected from a ResearchKit-based iPhone app from 1,756 families, finding that children with high autism risk evoked statistically significantly more neutral emotions and less positive emotions compared with children at low risk of autism [50].

Restrictive and repetitive head and body movements are another core trait defining autism, motivating work in analyzing quantitative differences in head movement, body movement, and other motor features between cases and controls. Martin et al. compared the yaw, pitch, and roll of the head of children with and without autism (21 children in each group), finding that children with autism exhibit faster head turning and inclination compared to neurotypical controls [51]. Dawson et al. conducted computer vision analysis of head postural control on 104 toddlers, 22 of whom had an autism diagnosis, finding that the rate of head movement for children with autism was 2.22 times greater than the neurotypical toddlers [52].

Another diagnostic criteria for autism is the use of idiosyncratic speech, giving rise to projects which analyze audio and text features. Hudenko et al., in a study of 15 children with autism and 30 typically developing children, found that children with autism produced no "unvoiced" laughter, which are largely atonal laughs without evident periodicity which are typically used to provide social affirmation rather than convey positive affect. In contrast, typically-developing children produced laughter which was 37-48% "unvoiced" [53]. Orlandi et al., in a study of 12 controls and 2 high rink infants, found differences in the mean fundamental frequency in the crying audio of high risk infants (siblings of children already diagnosed with autism) and controls at 10 days, 6 weeks, and 12 weeks [54].

Advances in human behavior analysis techniques are sure to bring about new analyses which will elucidate new behavior differences. As these methods continue to advance, incorporation with genomic and neural data will allow for an increasingly precise understanding of how small biological changes can affect behavior, helping to reveal which aspects of autism are preset biologically and which are affected by social and environmental factors. One of the largest opportunities for future case-control studies is a dramatic increase in scale. Given the diversity of humans, any study with tens of subjects in each cohort will contain a multitude of biases.

**Classification of Autism and Related Behaviors**

Supervised ML is one of the most widely applied data science techniques throughout most data-driven fields [55]. Naturally, the framework of training a model to make predictions from data can enable the creation of automated diagnostics, which can easily be formulated as a supervised learning problem where data from one of the modalities we discussed is used as the input to the model and a diagnosis is emitted as the model's output. We discuss research efforts following this general framework for autism and its comprising behaviors.

While automatic extraction of behaviors related to autism is feasible with current technologies (e.g., eye tracking for attention and facial emotion classification for affect), there are many behaviors which diagnosticians use to classify autism which are beyond the scope of current technologies. For example, "indicates pleasure to others", "shares excitement", and "social participation" are all diagnostic criteria for autism which appear as top predictors according to feature selection algorithms applied to clinical scoresheets [56]. For such behaviors, ML and other automated approaches are not feasible. Recently, ML-powered human-in-the-loop approaches to autism diagnosis have emerged. Such approaches either use untrained humans such as crowd workers [57-59] as human feature extractors of audiovisual data or direct in-person child observations such as parent-filled questionaries in conjunction with clinical experts and trained video analysts [60-64] as is the case with the pediatric digital health company Cognoa. After the human-extracted features are obtained, they are provided as input into an ML model, as in Tariq et al. who achieved an 92% AUC score [65] using 116 videos of children with autism and 46 videos of typically developing children. This pipeline was replicated in 50 neurotypical children and 50 children with autism from Bangladesh, achieving an AUC of 85% [66] and demonstrating that this method has the potential to generalize to diverse global populations.

An important design consideration when constructing diagnostic models is the identification of small feature subsets to minimize the time-to-diagnosis. Wall et al. pioneered in this application of feature selection methods to diagnostic scoresheets of autism [56]. Multiple works have subsequently validated the use of ML feature reduction methods to algorithmically identify small subsets of between 4 to 10 questions of diagnostic and screening instruments which alone can be used to predict autism [67-69].

Another important consideration for translation of these methods into real-world diagnostics is child privacy. Washington et al. applied privacy-preserving transformations to videos of children with autism including obfuscating the child's face with a box using frame-by-frame face detection, pitch shifting the audio [70], and applying global image transformations such as Gaussian blurring, pixelation, and optical flow [71]. Each of these mechanisms reduced performance of the human-in-the-loop models by less than 7% on metrics such as AUROC and AUPRC. There is ample room for future work to explore higher precision privacy mechanisms.

These human-in-the-loop diagnostic techniques have yet to be applied to behavioral conditions beyond autism, although we suspect that these methods could aid diagnostic practices for heterogeneous behavioral conditions more broadly, including attention-deficit/hyperactivity disorder (ADHD), anxiety, depression, and other developmental delays. We next discuss classification using such rich multimedia data streams. Although fully automated ML methods (as of 2023) cannot yet support the complexity of several social behaviors which encompass autism (e.g., *"Does the child enjoy participating in social games and interactions?"*), applying ML approaches to raw data streams for autism diagnostics has the benefit of picking up subtle cues and representations with speed and objectivity. We discuss these approaches, which leverage many of the same modalities as in the case-control studies.

Just as for case-control studies, gaze patterns are a common feature for diagnosing autism. Liu et al. used face scanning patterns to predict autism with an accuracy of 88.5% on a dataset of 29 children with autism, 29 neurotypical children matched by age, and 29 neurotypical children matched by IQ [72]. Duan et al. created a dataset of fixation maps and scanpaths of 14 children with autism and 14 controls [73], and datasets such as this can help support development of predictive models for autism. Chang et al. used a decision tree classifier to predict autism from a sample of 40 children with autism and 936 typically developing controls watching highly structured and strategically designed movies displayed on an iPhone or iPad with an AUC of 0.9 [74]. Oliveira et al. used gaze fixation maps from an eye tracker to predict autism on a dataset of 76 subjects with autism and 30 typically developing subjects, reaching 90% precision, 69% recall, and 93% specificity [75].

Affective computing is not as popular in automated diagnostic approaches as it is for case-control studies, although preliminary works exist. Drimalla et al. developed a tool to facilitate a simulated social interaction with a child, recording the subject's eye gaze, voice, and facial expressions during the interaction [76]. Using both facial expressions and vocal analysis collected from the simulated social interaction in a study with 37 adults with autism and 43 neurotypical controls, Drimalla et al. detected autism with an accuracy of 73%, specificity of 79%, and sensitivity of 67% [77].

Pose and movement features have been quite successful in distinguishing autism cases from controls. Khosla et al. used the arrangement of a child's eyes, nose, and lips in a front-facing image of the face to classify autism with 87% classification accuracy [78]. Hashemi et al. tracked head motion of children with autism using bounding boxes over the child's eyes, ears, and nose [79]. Lidstone et al. used 3D depth cameras from the Microsoft Kinect to distinguish children with autism from neurotypical controls, reaching an AUROC of 0.94 on a sample of 23 children with autism and 17 neurotypical controls [80]. Kojovic et al. achieved an F1 score of 0.89 in a binary autism diagnosis task on 68 children with autism and 68 neurotypical controls by using a CNN to extract visual features from pose images and feeding these into an LSTM network [81]. Cook et al. present a new dataset of video clips consisting of examples of 35 atypical motor movements associated with autism and 33 typical movements. They use this dataset to train a decision tree with skeletal keypoints extracted from OpenPose and temporal velocities encoding body movements over time, reaching an average F1 score across folds of 0.71 [82]. Li et al. classified autism using kinematic features extracted from hand movement imitations by 16 children with autism and matched controls, reaching an overall accuracy of 70.5% [83].

Anzulewicz et al. measured tablet touch and gesture kinematics on 37 children with autism and 45 typically-developing children, using these features to measure autism with 93% accuracy [84]. Cavallo et al. classified using hand movement data of 20 children with autism and 20 neurotypical children performing a structured grasping task, reaching 83% accuracy, 80% sensitivity, and 85% specificity [85].

Audio and text features are increasingly used for ML diagnostics of autism. Li et al. built a classifier to predict atypical prosody and stereotyped idiosyncratic speech associated with autism with accuracies of 88.1% and 77.8% respectively on a dataset of 118 children who were administered the ADOS Module 2 data and 71 children who were administered the ADOS Module 3 [86]. Chi et al. used crowdsourced videos from a mobile autism therapeutic game of 20 children with autism and 38 neurotypical children to classify autism with 79% balanced accuracy using a convolutional neural network predicting from spectrogram images [87]. Lau et al. used acoustic features pertaining to rhythmic and tonal aspects of prosody to classify autism achieved an AUC of 0.90 using rhythm-relevant features and 0.695 using intonation-relevant features on a dataset of 55 English-speaking people with autism and 39 English-speaking controls as well as 28 Cantonese-speaking people with autism and 24 Cantonese-speaking controls [88]. Maenner et al. constructed a random forest classifier of autism using words and phrases from developmental evaluations, observing 84% sensitivity, 89.4% positive predictive value, and an AUROC of 0.932 on a dataset of 1,450 children, of whom 754 met the criteria for an autism diagnosis [89].

Data modalities have recently been combined to predict autism. Vabalas et al. combined both eye and motion data to predict a diagnosis from 22 children with autism and 22 neurotypical controls, reaching 78% accuracy with both modalities, 73% for only motion features, and 70% for only eye features [90]. Javed et al. used both facial expressions and upper body movements recorded during child-robotic interactions to predict autism among 5 children with autism and 7 typically developing children with accuracy, precision, and recall in the high 80s [91].

Automated diagnostics for autism are sure to improve as ML methods continue to rapidly advance and as data collection systems become more ubiquitous. The classifiers will likely need continuous recalibration, as the definition of autism continues to evolve with subsequent iterations of the DSM. We anticipate a future where computerized diagnoses which are rapid and free to use will become widely available globally.

**Translation of Diagnostic Models**

The vibrant research in the space of artificial intelligence for automatic evaluation and measurement of observable phenotypes of autism (and other) has promise to translate, likely sooner than later, to practical clinical and educational solutions that enable diagnosis, tracking, and treatment. This requires careful considerations of design, feasibility testing, and tests for clinical effectiveness using useful endpoints. In this section we focus just on the diagnostic and therapeutic tools that have passed through regulatory review and have received market clearance.

Diagnostic tools for autism have started to penetrate the market. Cognoa [60-64] has an FDA-approved digital diagnostic which uses artificial intelligence for clinical decision support in the identification and ruling out of autism in children ages 18-72 months called Canvas Dx. This diagnostic device is indicated for remote use on smartphones as well as for use in both primary settings and more specialty settings. Another diagnostic tool, Earlitec [40, 92-98], incorporates the single modality of eye tracking for autism diagnosis support in children ages 16-30 months old and is approved for use specifically in the specialty clinical settings.

Several have noted the importance of managing uncertainty with AI in medicine [99-101]. Due to the importance of precision measurement when making AI-guided diagnoses on developing children, where ephemeral signals are common and expected, managing uncertainty is particularly important. Cognoa's canvasDx [60-64] is one such tool. CanvasDx emits a probability score of autism but only makes a positive call of "autism" or a negative call of "no autism" if the emitted probability exceeds an upper and lower threshold, respectively. Otherwise, the classifier abstains from classification, indicating that the model is not confident enough to make a prediction. The ability to abstain allows the model to avoid making diagnostic calls when presented with insufficient information to make a call in either direction. This is an important safety measure, but one that can be tuned and optimized with new information, for example under the FDA's recent guidance for incorporation of a predetermined change control plan, which has been granted with the authorization of de novo and derivative devices including CanvasDx, Medtronic, Caption Guidance, and Apple's A-fib detection system. We anticipate that all medical devices using AI for diagnostics will be required by the FDA to have an abstention feature and that many will include algorithm modification plans that enable careful tuning with good machine learning practices given real word use data.

Evaluating these diagnostic models which can abstain from making a prediction requires careful consideration of evaluation metrics. We argue that positive predictive value (PPV) and negative predictive value (NPV) are the optimal evaluation metrics for models which make abstentions, as they only consider the positive and negative calls which are made rather than metrics like sensitivity and specificity which consider all positives and negatives regardless of whether the classifier made a call on those cases. We therefore recommend using PPV and NPV as the primary clinical endpoints in diagnostic classifiers.

Another crucial endpoint for diagnostic systems is model fairness. A common issue with ML systems for healthcare is that they have differing performance levels across demographic groups [102-103]. There are several quantitative fairness metrics which have been proposed in recent ML literature, including demographic parity, equalized odds, equal opportunity, test fairness, and counterfactual fairness, among many others [103]. While no fairness metrics have been standardized yet in the field, we believe that moving forward, it will be important to report standard metrics like demographic parity along with standard performance metrics. Large crowd data collection efforts such as Simons Foundation Powering Autism Research for Knowledge (SPARK) [104] and remotely and freely accessible smartphone applications [20-23] are important steps towards achieving fair and unbiased data collection.

**Translation of Models for Digital Therapeutics**

There is tremendous promise for digital solutions that use ML to improve treatments delivery and quality for autism and other aspects of behavioral health. Because AI-powered digital therapy is a blossoming field of study, there are uncertainties about the reactions of patient populations to the therapies on dimensions of trustworthiness, usability, and understandability, among several others.

We describe two AI-enabled solutions for pediatric autism which have followed this path to translation. The SuperpowerGlass system is one such therapy which is delivered on a Google Glass. In therapy via SuperpowerGlass, the parent interacts with the child with autism while the child wears the Glass system. The Glass provides real-time emotion cues to the child via a computer vision facial emotion detection system. SuperpowerGlass was developed and refined through an iterative design process. In early prototyping sessions, possible user interface designs were explored through storyboarding and low-fidelity prototyping with children with autism and their parents through co-design sessions [32-35]. These changes were made prior to initial feasibility testing [36-37], which led to further refinements until the system was formally tested in a randomized controlled clinical trial [38]. This process is crucial for novel therapeutic paradigms in digital health given the uncertainty and novelty of AI-powered solutions.

Another therapy, EndeavorRx by Akili, is the first game-based digital therapy and one of the first digital therapies to pass through regulatory review [105]. This system is a sophisticated design with underlying software models for independent game play by a child with ADHD, a condition that has numerous symptomatic overlaps with and is often co-morbid with autism. Both EndeavorRx and SuperpowerGlass have translated to market and have received "Breakthrough Therapy" status by the FDA.

Digital applications for psychiatry and behavioral sciences are traditionally developed as "one-size-fits-all" solutions. However, the integration of the machine learning models discussed in this review can enable adaptive digital tools which analyze the user's behavior and recalibrate accordingly. Measures of autism symptoms can be embedded within gamified therapeutics so that the AI can understand the child's response to a particular "dose" of the therapy in real time. With such an understanding of the child's dose-response curve, digital solutions can then increase or decrease the dose of the therapy.

**Challenges and Opportunities**

The field of autism behavioral data science is a special case of computational behavioral phenotyping. As such, the field is likely to continue improving as computational behavior analysis tools become more sophisticated, granular, and robust. Similarly, autism data science can benefit from advancements in behavioral data science efforts for other psychiatric conditions. We briefly discuss challenges and opportunities in the field of digital behavioral phenotyping for neuropsychiatry. Many of the opportunities we discuss involve reapplying data science innovations used in other conditions to autism, and many of the challenges for autism data science also face data science for complex human behavior analysis more broadly.

One of the most important difficulties in the field of autism data science is the lack of large, standardized datasets reported in studies. Classification performances reported by ML diagnostic systems vary drastically across papers, but it is unclear whether these performance differences are a result of datasets with intrinsically varying differences. There is therefore a need for standardized datasets. Similarly, the sample sizes in both case-control and ML analyses are usually small, as there tend to be only tens of children represented in each group. The reported numbers from these studies limit the ability to compare across studies even if the same study conditions are used. There is a strong need for larger and more diverse studies, as these will enable generalizability of findings. Crowdsourced domain-specific data collection platforms [20-23] hold great promise for enabling these sample sizes. However, unlike controlled lab settings, such systems face issues of data consistency, resulting in severe overfitting. An alternative approach which bypasses this issue is model personalization: the model is optimized for each subject's unique features and the environment in which the model will be deployed. There are some preliminary efforts published in the autism data science field [106] which demonstrate improved performance when personalizing ML classifiers to the subject. These approaches require increased interaction between the child and the subject, creating opportunity for novel human-computer interactions.

Self-supervised learning (SSL) is the automatic learning of feature representations from unlabeled data. SSL can help accelerate model personalization when applied to a single user's data streams. Yu and Sano applied SSL to wearable sensor data for stress detection, improving performance compared to supervised baselines by between 7.7% to 13.8% on 3 independent datasets [107].

Distinguishing autism from related conditions is an emerging challenge for the field. An increasing body of literature from autism genetics, neuroscience, and psychiatry suggest that autism does not consist of a single spectrum but instead as independent subtypes. There have been some efforts to quantitatively subtype autism from behavioral data modalities, and there is an opportunity for such efforts to validate and complement efforts in genetics, neuroscience, and psychiatry. For example, Gardner-Hoag et al. analyzed 854 children with autism and identified 7 distinct clusters of autism behaviors, with each cluster representing a single dominant prototypical behavioral characteristic [108].

While most of the analyses reported here distinguish between having autism and not having autism or being neurotypical, few efforts to date have attempted to distinguish between a variety of psychiatric conditions with overlapping symptoms. Binary classifiers are useful if a condition is already suspected, but the ideal scenario will be to distinguish autism from conditions such as ADHD, schizophrenia, anxiety, depression, and speech delays. There has been some preliminary work in this vein. Duda et al. built a classifier distinguishing ADHD (N=174) from autism (N=248) using a subset of 15 questions from the SRS, reaching an AUROC of 0.89 [109]. Wawer et al. used text utterances to classify both autism (N=37) and schizophrenia (N=37) separately, finding that pretraining a model for schizophrenia results in improved performance when transfer learning to predict autism [110]. Demetriou et al. used a battery of tests created for various psychiatric conditions to classify autism (N=62), early psychosis (N=48), and social anxiety disorder (N=83), suggesting that combining inputs for different overlapping conditions can be a useful approach [111]. Iakovidou et al. used accelerometer, blood volume pulse, and electrodermal activity data from wearable sensors to differentiate autism from Rett's syndrome with 95% accuracy in a sample containing 10 children with autism and 10 with Rett's syndrome [112]. Since neuropsychiatric conditions are often comorbid, future efforts will likely need to be formulated as a multiclass classification problem or a similar variant.

In addition to case-cohort analyses and ML diagnostic systems, data science methods have the potential to provide automated longitudinal outcome tracking and treatment response measurements. Some preliminary work in this area has been conducted, but there are ample opportunities for more work in this field, which would benefit digital interventions and treatments for autism symptoms. McKernan et al. measured the amount of time for a child to respond to a social partner using automated speech processing methods, finding that children receiving interventions demonstrated a larger reduction in latency compared to those not receiving the intervention (105 children in each group) [113]. Kolakowska et al. measured tablet sensors during gameplay, finding that these sensors can be used to predict therapy progress in 40 children with autism with accuracy surpassing 80% [114].

The study of autism spans several academic disciplines, and there are myriad data science studies for autism using genomics [115-118], epigenomics [119], proteomics [120-123], metagenomics [124-126], and brain imaging [127-130]. While we admire these works, these studies are beyond the scope of this review. We note, however, that there is an opportunity for efforts in biology-based fields to eventually converge with the types of analyses describe here, as increasingly granular behavioral phenotypes using digital phenotyping methods will likely enable more robust and precise biological association studies. We note that many of the data science methods which we have reviewed can apply more broadly to neurological, developmental, and behavioral conditions.

We invite all readers to join us in this promising, burgeoning field. We look forward to the increases in access to care and scientific understanding of complex social human behavior which will inevitably flourish over the next two decades and beyond.